\documentclass[twocolumn,trackchanges]{aastex61}

\usepackage{soul}

\usepackage{comment}

\newcommand\aastex{AAS\TeX}

\submitjournal{ApJ}

\newcommand\lya{Ly$\alpha$}
\newcommand\msun{M_\odot}
\newcommand\mkms{\rm km \, s^{-1}}

\shorttitle{\aastex\ CGM of LRGs}
\shortauthors{Smailagi{\' c} et al.}

\begin{document}

\title{Extreme circumgalactic H~\textsc{i} and C~\textsc{iii} absorption 
around the most massive, quenched galaxies}

\author{Marijana Smailagi{\' c}}
\affil{University of California, 1156 High Street, Santa Cruz, CA 95064, USA \\
}
\email{msmailag@ucsc.edu}

\author{Jason Xavier Prochaska}
\affiliation{University of California, 1156 High Street, Santa Cruz, CA 95064, USA \\
}

\author{Joseph Burchett}
\affiliation{University of California, 1156 High Street, Santa Cruz, CA 95064, USA \\
}

\author{Guangtun Zhu}
\affiliation{Johns Hopkins University, 3400 N. Charles Street, Baltimore, MD 21218, USA \\
}

\author{Brice M{\' e}nard}
\affiliation{Johns Hopkins University, 3400 N. Charles Street, Baltimore, MD 21218, USA \\
}

\begin{abstract}

Luminous red galaxies (LRGs) are the most massive galaxies at $z\sim 0.5$ and, by selection, have negligible star formation. 
These objects have halo masses between those of $L_{*}$ galaxies, whose circumgalactic media (CGM) are observed to have 
large masses of cold gas, and clusters of galaxies, which primarily contain hot gas. 
Here we report detections of strong and extended metal (\ion{C}{3} 977) and 
\ion{H}{1} lines in the CGM of two LRGs.  
The \ion{C}{3} lines have equivalent widths of $\sim 1.8$ \AA\ and $\sim 1.2$ \AA , and velocity 
spreads of $\sim 796~\mkms$ and $\sim 1245~\mkms$,
exceeding all such measurements from local $\sim L_{*}$ galaxies (maximal \ion{C}{3} equivalent widths $\sim 1$ \AA). 
The data demonstrate that a subset of halos hosting
very massive, quenched galaxies contain significant complexes of cold gas. 
Possible scenarios to explain our observations include that the LRGs’ CGM originate from past 
activity (e.g.,\ star formation or active galactic nuclei driven outflows) 
or from the CGM of galaxies in overlapping subhalos.
We favor the latter scenario, in 
which the properties of the CGM are more tightly linked to the underlying dark matter halo than properties 
of the targeted galaxies (e.g., star formation).

\end{abstract}


 \section{Introduction} \label{sec:intro}

A series of absorption-line surveys piercing the halos of individual
galaxies have established the near-ubiquitous presence of a 
large reservoir of cool and enriched gas in the circumgalactic
environment.  This cool circumgalactic
medium (CGM) is manifest in galaxies with a wide range of luminosity
and across most of cosmic time
\citep[e.g.,][]{Tumlinson17,Chen17}. 
In the `normal' population of field galaxies, one now recognizes
the cool CGM as a fundamental baryonic component of galaxies and,
therefore, a main cog in the machinery of galaxy formation. 

From this ensemble of observational CGM studies, several key trends and
puzzles emerge.  Of particular interest to this work is the apparent
correlation between the absorption strength of the cool CGM
and galaxy mass \citep{Chen10,X11,X14}.
Most recently, \cite{Bordoloi17} has emphasized a strong correlation
between the \lya\ equivalent width (EW) and the galaxy stellar mass in present-day,
$L \lesssim L^*$ galaxies.  These authors interpret the increase in 
EW
with stellar mass as a fundamental relation between the CGM dynamics and the underlying dark matter halo mass.
Another, more puzzling, result that has materialized is the preponderance of 
this cool CGM even around quiescent galaxies 
\citep{Thom12,Zhu14,Huang16,Zahedy16,Tumlinson17,Chen17}, 
but see also \citet{Lan14}. 
Despite an interstellar medium 
in many cases
nearly devoid of cool gas and minimal active star-formation 
\citep[e.g.,][]{Young14}, 
these galaxies exhibit CGM
with HI and low-ionization metal content 
comparable to the star-forming population. 
Extrapolating these two results to higher mass halos, where red
and dead galaxies proliferate, one might conclude that 
galaxy clusters may harbor the largest mass of cool halo gas in the
universe.

This assertion, however, runs contrary to observations of the hot
intracluster medium in galaxy clusters \citep{Mitchell76} 
and theoretical work which predicts a predominantly virialized gas
\citep{Voit05,Dekel06,Kravtsov12}. 
And, indeed, the few studies that
have examined the cool CGM of galaxy clusters support a 
suppressed incidence of such gas both within the halo and within
the halos of the cluster members 
\citep{Burchett17,Yoon17}, 
but see also \citet{Lopez08}.
These results, while still sparse, suggest a rapid decline in 
the cool CGM in the most massive halos.

Another opportunity to extend CGM studies into the high 
halo-mass regime is afforded by the set of over 1 million luminous
red galaxies (LRGs) discovered in the Sloan Digital Sky Survey
\citep{Dawson13}.  These galaxies have inferred halo masses $\approx 10^{13.5} \msun$ \citep{White11}  
and, by selection, have negligible on-going star formation.
Focusing on the \ion{Mg}{2} doublet (the best cool CGM diagnostic
available at optical wavelengths for intermediate redshifts), CGM 
surveys of LRGs have demonstrated a lower incidence of such gas than around $L^*$ galaxies, with covering fraction of $\sim 5$ \%  
\citep[e.g.,][]{Huang16}.
Nevertheless, the full statistical power of the BOSS experiment
has revealed a strong excess in the cross-correlation
between \ion{Mg}{2} and the LRG within the virial radius
\citep[][and references therein]{Zhu14}
Furthermore, the relation of estimated mean 
EW
with impact parameter approaches that of $L^*$ galaxies
\citep[][fig. 10]{X14}.

Inspired by the \cite{Zhu14} survey, we pursued far-UV observations
for a modest sample (15) of QSO sightlines probing the LRG-CGM.
Our data access the strong \ion{H}{1} Lyman series and
metal-line transitions that trace intermediate (e.g.\ \ion{C}{3})
and highly ionized gas (\ion{O}{6}) at the LRG redshifts, which 
ground-based surveys did not.
In this Letter, we report on the surprising detection of two LRGs whose
cool CGM absorption exceeds that of all previous measurements of 
$L \sim L^*$ galaxies.  
This manuscript is organized as follows:
In Section \ref{sec:data}, we briefly describe the data and spectral line measurements.
Section \ref{sec:dis} puts these two LRGs in the broader context 
of CGM observations and discusses the origin of our
extreme CGM cases.

\section{Data, sample and measurements} \label{sec:data}

The two LRG-QSO pairs analyzed here
are taken from a larger survey studying the CGM of LRGs, 
which we describe here briefly. 
The SDSS-III/Baryon Oscillation Spectroscopic Survey 
\citep[BOSS;][]{Dawson13} 
contains over $10^{6}$ spectra of LRGs at redshifts $z\sim 0.5$.
We matched LRGs from the BOSS survey and QSOs from 
the SDSS DR12 survey  \citep{Paris17}, 
such that the projected distance between the QSO and the LRG (impact parameter) 
is less than $400$ kpc, and the LRG has a redshift $z_{\rm LRG} > 0.4$. 
From this sample we selected 15 pairs 
where the quasars have
bright GALEX magnitudes and 
where
approximately half 
were 
previously 
known to exhibit Mg\textsc{ii} absorption
in the LRG-CGM \citep{Zhu14}.  
A complete survey description is forthcoming by
Smailagi{\' c} et al. (2018, in preparation).

\begin{figure*}[t!]
\centering
\includegraphics[width=0.95\textwidth]{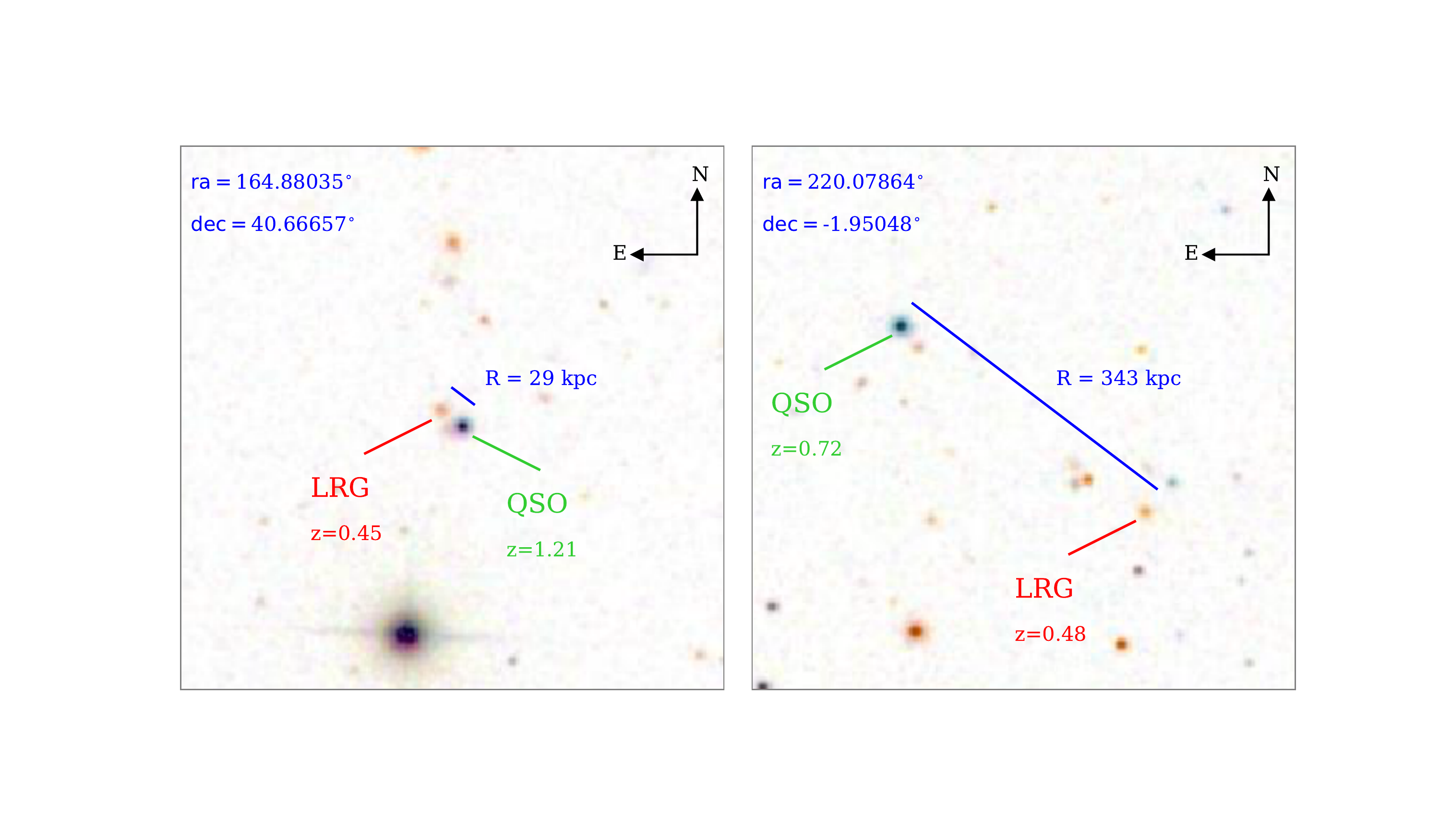}
\vskip -0.5in
\includegraphics[width=0.55\textwidth]{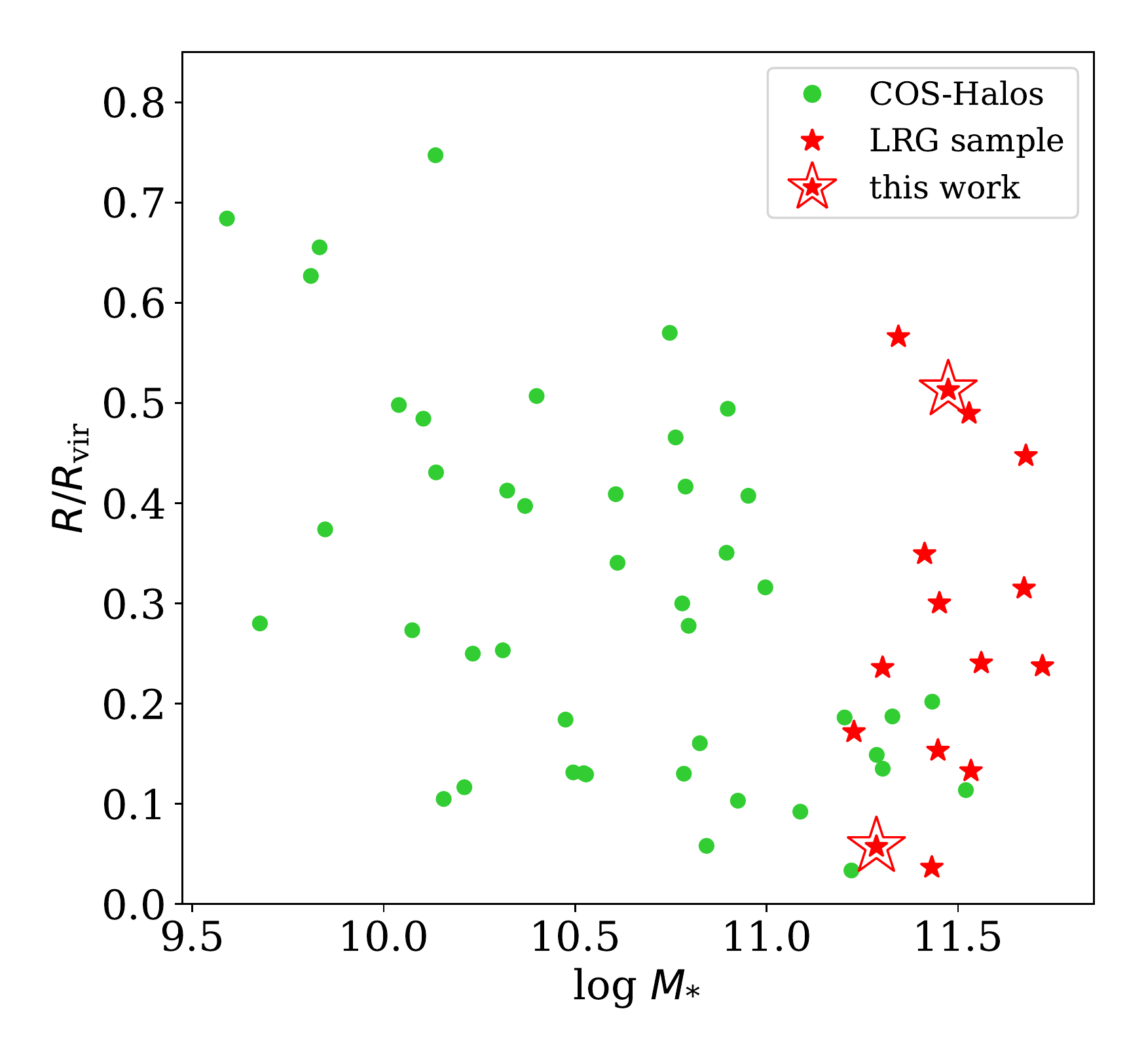} 
\vskip 0.in
\caption{\small 
Top: SDSS images of the two LRGs. 
The LRG coordinates are given in the upper left corner.
Bottom: Impact parameter $R$ 
scaled to the virial radius $R_{\rm vir}$ 
versus
stellar mass $M_{*}$
for LRGs and COS-Halos. 
}
\label{fig:lrgs}
\end{figure*}

Images from the Sloan Digital Sky Survey (SDSS) of 
LRG\_1059+4039 and LRG\_1440-0156
are shown in Figure~\ref{fig:lrgs}. 
Both LRGs are located at $z\sim 0.5$
and have impact parameters of $\sim 30$ kpc and $\sim 350$ kpc, 
typical of the full sample.  
The bottom panel in Figure \ref{fig:lrgs} shows the impact parameter 
(scaled to the virial radius\footnote{
We estimate halo masses from stellar masses and 
\citet{Moster10} relation between stellar and dark matter halo mass.
Stellar masses are used from the Wisconsin group,
wisconsin\_pca\_m11-DR12-boss.fits.gz 
\url{http://www.sdss.org/dr14/spectro/galaxy_wisconsin/} )
Virial radius is calculated as for COS-Halos in \citet{Tumlinson13}.
}) versus 
the stellar mass $M_{*}$ for our full sample 
of 15 LRGs; galaxies from the COS-Halos survey 
\citep{Werk12}
are also plotted for reference. 

We obtained HST COS spectra of the 15 QSOs
in Cycle 23 (GO-14171 ; PI Guangtun Zhu). 
We used the G140L grating with 
resolving power $R \sim 2000 - 3000$ for the wavelengths of
interest 
(spectral resolution $\sim 150$ km s$^{-1}$). Wavelength coverage for the segment A is $\sim 1300 - 2000$ \AA, and signal-to-noise ratio per pixel is $\sim 4-6$.
Data were reduced using the CALCOS v2.21 software package
and our own custom Python codes\footnote{\url{https://github.com/PYPIT/COS_REDUX}}.
Our modifications to the CALCOS software 
follow those described by \citet{Worseck16}, and include a modified estimate 
of the dark current and modified co-addition of sub-exposures. 
In the reduced spectra, the quasar continuum was 
estimated by generating a cubic spline `anchor' points set by eye using
the lt\_continuumfit GUI of the {\sc linetools} 
package\footnote{\url{https://github.com/linetools/linetools}}.
Absorption lines were identified with the {\it igmguesses} 
tool from the {\sc PYIGM}  package\footnote{\url{https://github.com/pyigm/pyigm}}. 
EWs
were calculated from box-car integration across the absorption complex.
Line detections are considered reliable if their 
EWs
exceed three times their estimated uncertainties.

\begin{figure*}[t!]   
\centering
\includegraphics[width=0.45\textwidth]{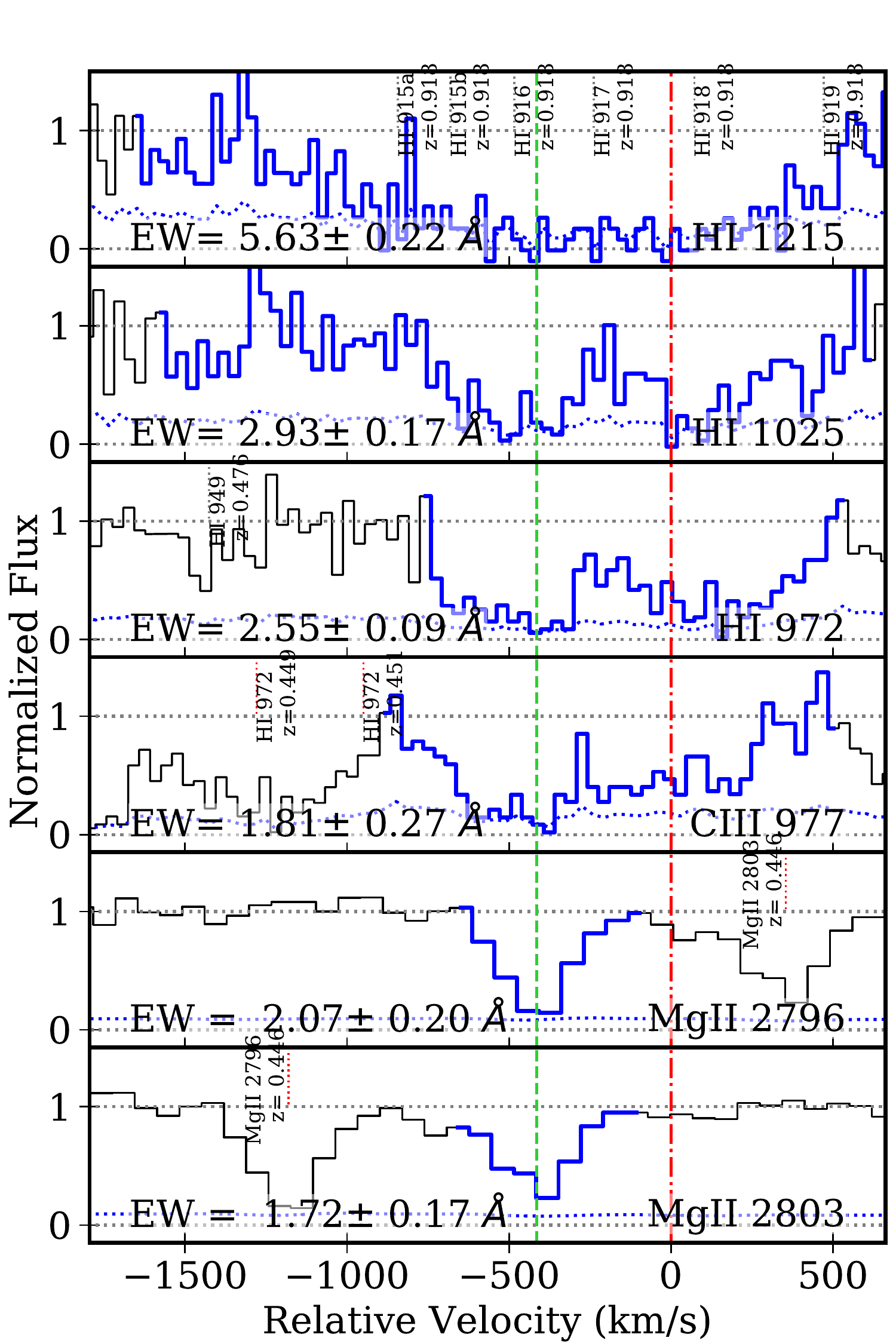} 
\hspace{0.2in}
\includegraphics[width=0.45\textwidth]{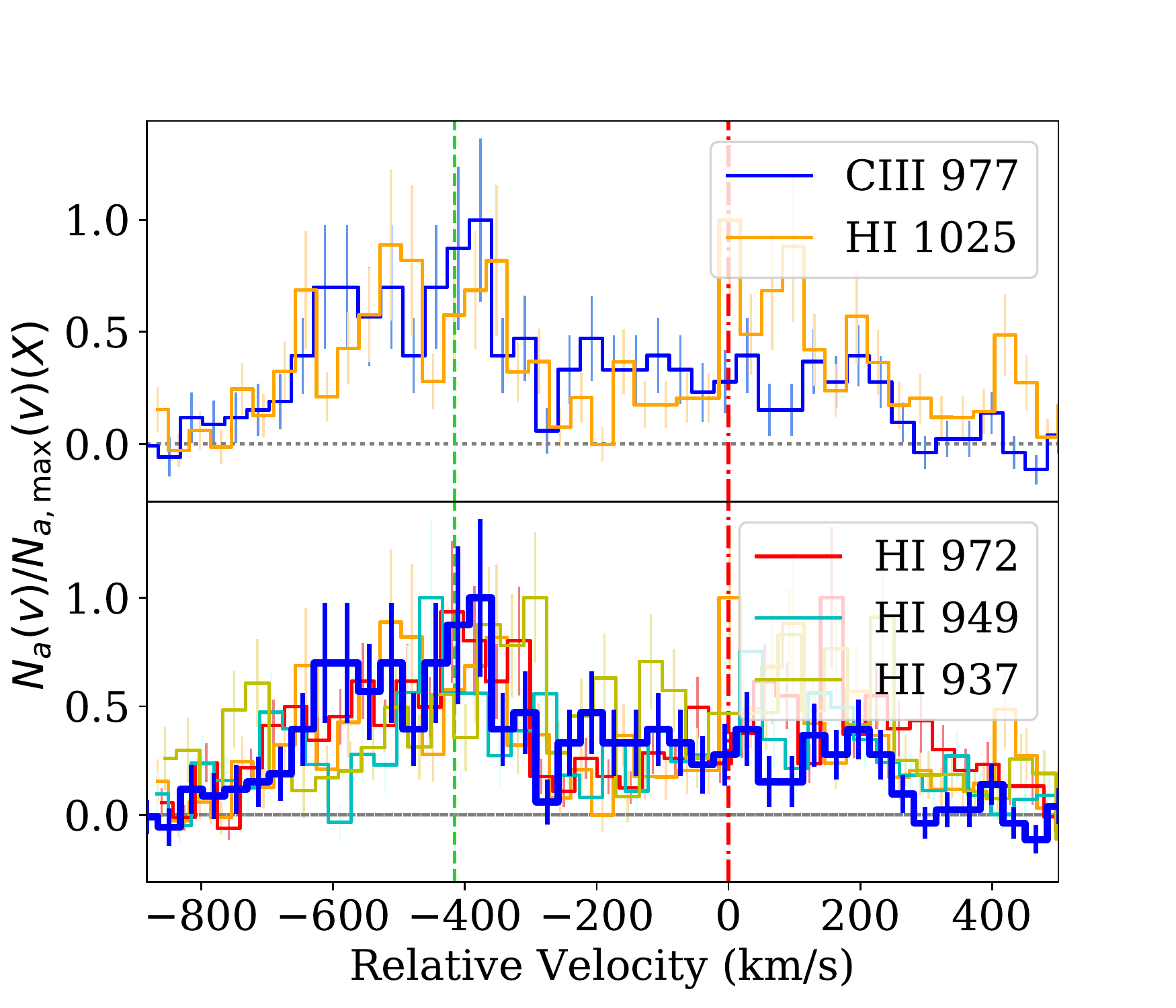}
\vskip 0.in
\caption{\small 
Left:
Line profiles for LRG\_1059+4039: H~\textsc{i} 1215,
H~\textsc{i} 1025,
H~\textsc{i} 972, C~\textsc{iii} 977, Mg~\textsc{ii} 2796, and 
Mg~\textsc{ii} 2803.
The analyzed velocity ranges are marked in blue. Interloping transitions in each panel are labeled in red if they originate from the same LRG CGM system, otherwise in gray. Rest frame equivalent widths are shown in the lower left corner. The dotted blue line represents the flux uncertainty. 
Right: Apparent column densities of the \ion{C}{3} and \ion{H}{1} transitions associated with LRG\_1059+4039, scaled such that the maximum of each is unity.
Velocities at the redshifts of the LRG and Mg~\textsc{ii} absorption are marked with red dot-dashed and green dashed vertical line, respectively. 
}
\label{fig:lines}
\end{figure*}

\begin{figure*}[t!]  
\centering
\includegraphics[width=0.45\textwidth]{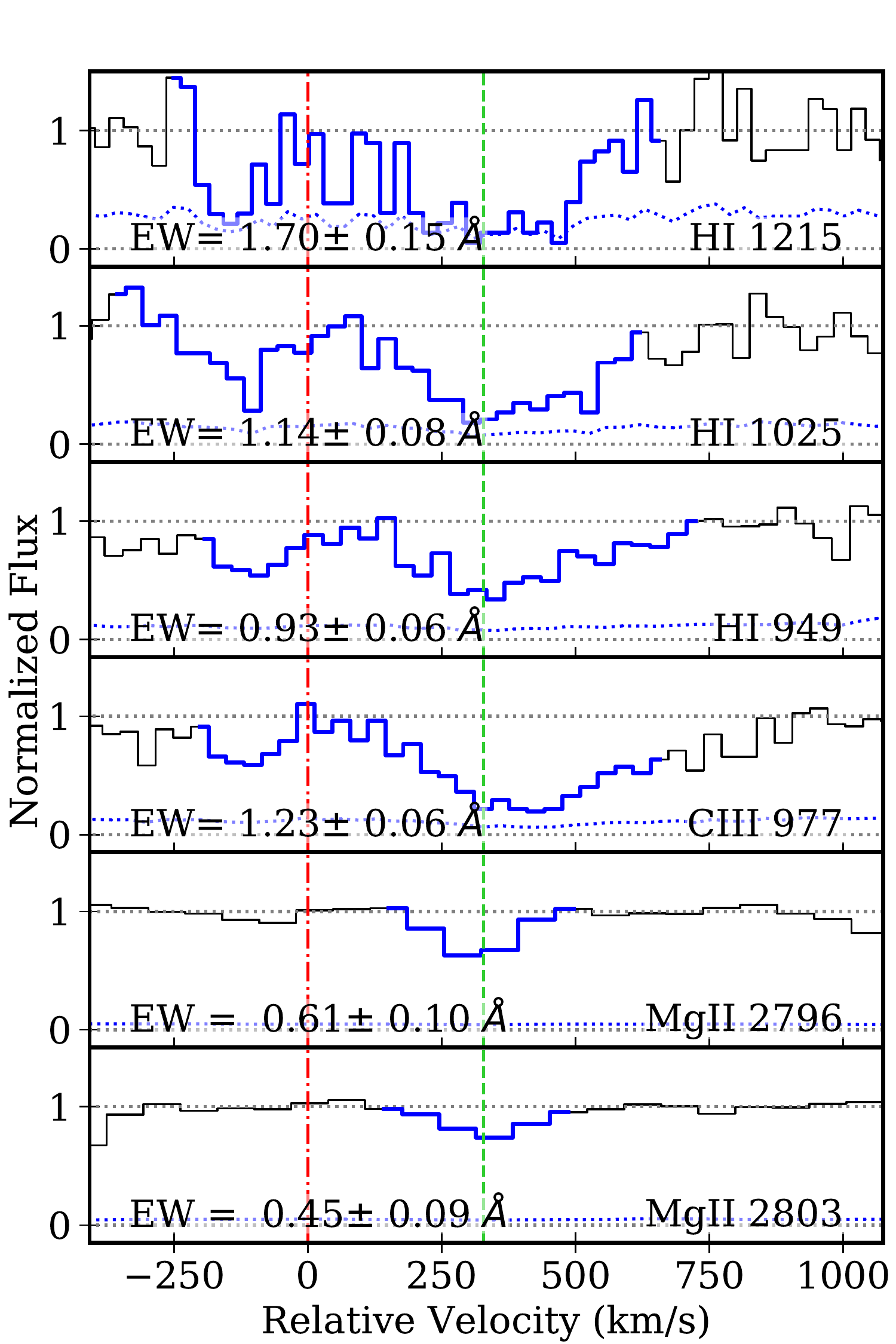} 
\hspace{0.2in}
\includegraphics[width=0.45\textwidth]{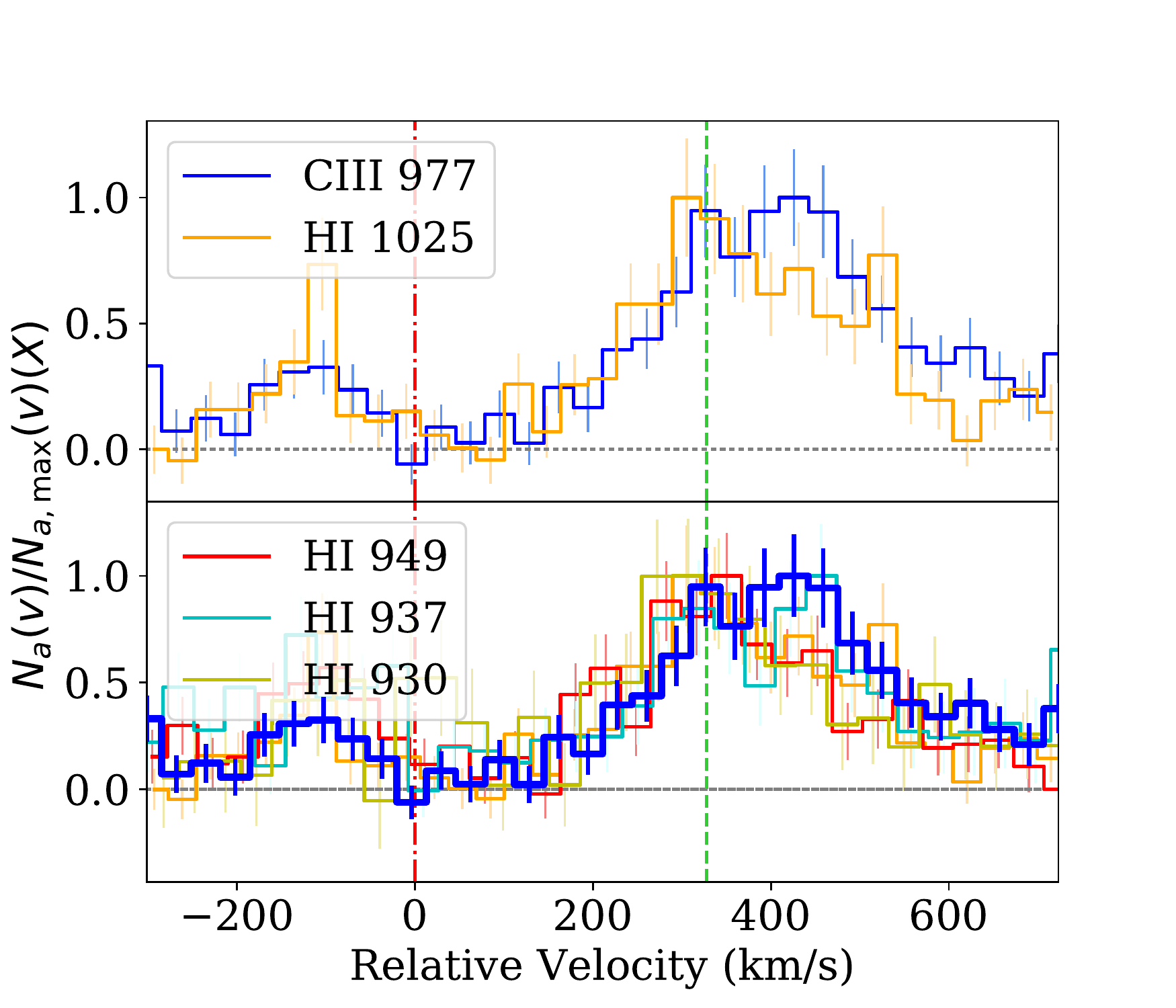}
\vskip 0.in
\caption{\small 
Same as Figure \ref{fig:lines}, but for LRG\_1440-0156. 
}
\label{fig:lines2}
\end{figure*}

Regarding line-identification, which is non-trivial for
sightlines to $z \sim 1$ quasars,
we first identified strong lines in the Milky Way and those associated with the QSO. 
We then searched for lines near the LRG redshift, beginning with the 
\ion{H}{1} Lyman series and proceeding to metal-line transitions.
We also searched for unrelated, strong absorption systems along the sightline
that could present 
interloping absorption with the LRG-CGM lines.
A subset of the transitions identified with the LRG-CGM is shown
in  Figures~\ref{fig:lines} and \ref{fig:lines2}. 
It is evident that these galaxies exhibit very strong low, intermediate, and even high-ion absorption.
One also notes that the strongest low and intermediate-ion absorption is coincident with the largest optical depth of \ion{Mg}{2} from the SDSS QSO spectra.

To further confirm the line identifications, 
Figures \ref{fig:lines} and \ref{fig:lines2} also compare
scaled apparent column densities of C~\textsc{iii} 977 and 
H~\textsc{i} transitions. For example,
from Figure \ref{fig:lines2} 
we see that the
scaled apparent column density profiles for C~\textsc{iii} 977 and H~\textsc{i} 1025 are consistent 
in the velocity range from $\sim$ -300 to 400 km s$^{-1}$. 
At velocities $> 400$ km s$^{-1}$, the \ion{C}{3} line is possibly blended, or the physical conditions of the gas may differ, changing the \ion{C}{3}/\ion{H}{1} ratio as well.
This close alignment between the absorption from these two species over such a large velocity range corroborates the identification of this system while revealing an
extraordinary complex of metal-enriched gas in the CGM of the LRG.
We find similar results for LRG\_1059+4039 
and for the other H~\textsc{i} transitions.

Next, we measure the velocity spread $\Delta v_{90}$ for \ion{C}{3} 977, which is defined as the velocity interval containing 90\% of the total optical depth \citep{PW97}. 
Pixels where flux is below noise level were replaced with the noise.
The two LRGs have $\Delta v_{90} \sim 945$ and $\sim 760$
km s$^{-1}$, respectively. 
These values will decrease by $\sim 100$ $\mkms$ if consider the COS spectrograph resolution. We also caution that the \ion{C}{3} 977 lines are saturated. 
The Ly$\alpha$ lines are also saturated, but we obtain $\Delta v_{90, {\rm Ly}\alpha}$ = 1497 and 640 $\mkms$ if we measure them in the same manner.

\section{Results and Discussion} \label{sec:dis}

\begin{figure}[h]
\centering
\includegraphics[width=0.49\textwidth]{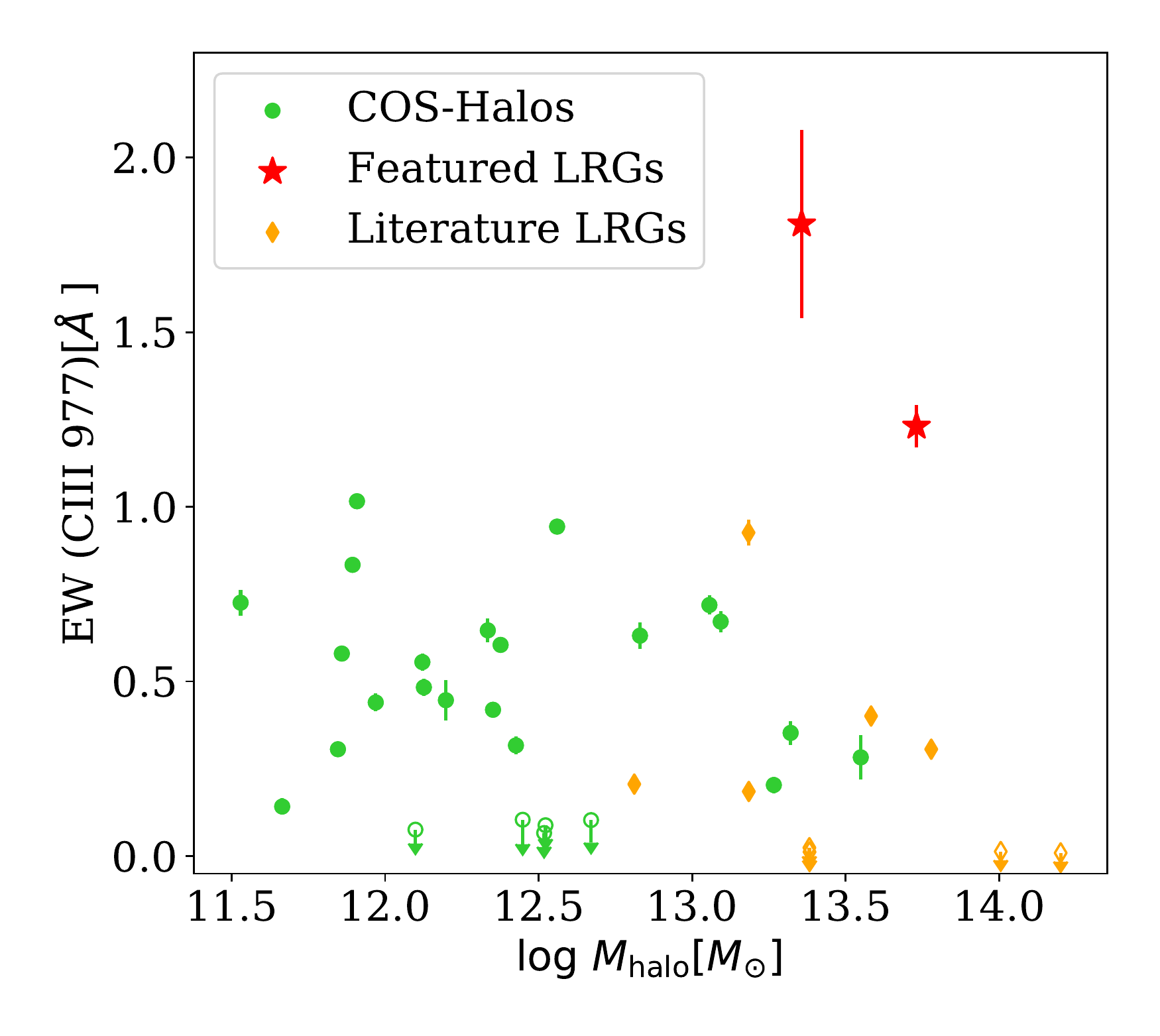}  
\vskip -0.16in
\caption{{\small 
Rest frame EW of the C~\textsc{iii} 977 \AA\ line versus dark matter halo mass for our two LRGs (red stars) compared with COS-Halos (green circles) and literature LRGs \citep[orange diamonds;][]{Chen18}$^{a}$ 
EWs for the two LRGs exceed those of all COS-Halos galaxies, and the corresponding EW for LRG\_1059+4039 is nearly twice the COS-Halos maximum.  }
\newline
{\footnotesize{$^a$ We found that in \citet{Chen18} LRGs are defined differently than as originally defined in BOSS survey (see e.g. LRGs coordinates list in \url{http://www.sdss.org/dr14/spectro/galaxy_wisconsin/}), 
and we display only their LRGs that were initially defined as such.} }}
\label{fig:ewmst}
\end{figure}

\begin{figure}[h]
\centering
\includegraphics[width=0.49\textwidth]{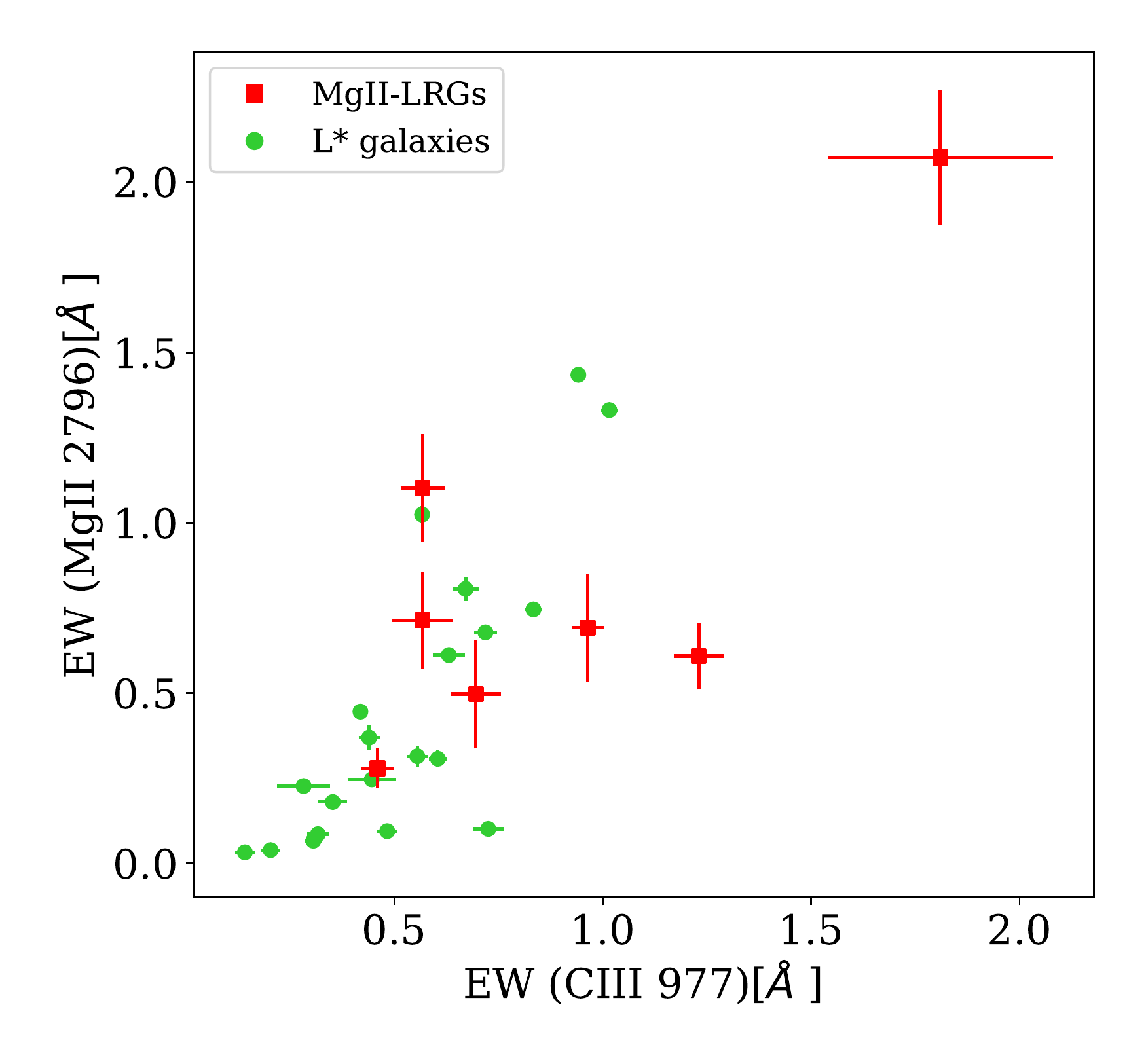}  
\vskip -0.16in
\caption{\small 
Equivalent width in \ion{Mg}{2} 2796 versus equivalent width in \ion{C}{3} 977 for COS-Halos (green circles) and LRGs from our full sample (red squares).}
\label{fig:ciiimgii}
\end{figure}

The spectra shown in Figures~\ref{fig:lines} and
\ref{fig:lines2} demonstrate the presence of extremely
strong, cool CGM absorption-profiles surrounding LRG\_1059+4039 and LRG\_1440-0156.
To provide context, we compare the measured \ion{C}{3}~977 
EWs
against the data for low-$z$ $L^*$ galaxies
from the COS-Halos survey \citep[Figure~\ref{fig:ewmst};][]{Werk13}.
The C~\textsc{iii} 
EWs
of the two LRG CGM 
($\sim 1.8$ \AA\ and $\sim 1.2$ \AA\ )
clearly exceed the entire $L^*$ galaxy distribution.
In fact, the LRG\_1059+4039 \ion{C}{3} absorption is 
nearly twice as strong as the $L^*$ CGM maximum.
To date, the only other population of galaxies known to exhibit such
strong and extended absorption lines are QSO host galaxies 
at $z \sim 2$ \citep{X13,Lau16} 
and ultra-strong \ion{Mg}{2} absorbers \citep[e.g.,][]{Nestor11}.
We note that a few other individual examples of large low-ion 
EWs
associated with 
the CGM of
luminous red 
or massive elliptical
galaxies \citep{Gauthier13,Zahedy16} have been reported.  
However,
none of these studies found metal UV transitions exceeding those of $L_{*}$ galaxies.
Figure~\ref{fig:ewmst} also shows that
our two LRGs and some of the $L_{*}$ galaxies exceed 
other known LRGs CGM with available \ion{C}{3} measurements \citep[recently studied by][]{Chen18}.
We note that certain $L^{*}$ galaxies are associated with ultra-strong \ion{Mg}{2} absorbers \citep[e.g.,][]{Nestor11}; however, these associations are rare and are possibly (post-)starbursts. Nevertheless, \ion{C}{3} measurements are not available for these strong absorbers.
Using the combined sample of COS-Halos \citep{Werk13} and 
our sample of LRGs, a Spearman correlation test shows a $\sim 5$-sigma correlation between EW(\ion{C}{3}) and EW(\ion{Mg}{2}), see Figure~\ref{fig:ciiimgii}. 
This indicates that \ion{C}{3} and \ion{Mg}{2} might track one another.
However, the scatter in this relation is $\sim 1$ \AA, 
and if we fit a linear relation between EW(\ion{C}{3}) and EW(\ion{Mg}{2}), then for 3 of 7 LRGs, the difference between the measured and predicted EW(\ion{C}{3}) will be greater than 0.4 \AA\ (for all COS-Halos systems, the differences are $\lesssim 0.4$ \AA ). 
For example, one of the large-EW(\ion{C}{3}) LRGs featured in this work shows an EW(\ion{Mg}{2}) that lies $\sim 1$ \AA\ below this would-be relation and in fact has a smaller EW(\ion{Mg}{2}) than several of the EW(\ion{C}{3}) $< 1$ \AA\ COS-Halos and LRG systems.  
In addition, the same correlation for absorbers with EW(\ion{C}{3}) $> 0.5$ \AA\ (LRGs from our full sample, COS-Halos, \citet{Chen18} galaxies) is significant to only 1.1 $\sigma$. 
Given this and the large scatter in a such a correlation, we contend that \ion{Mg}{2} and 
  \ion{C}{3} offer independent information on the CGM of LRGs.  Certainly, the presence of \ion{Mg}{2} absorption does not ensure such strong and widely extended \ion{C}{3} as we report here (cf. the \ion{Mg}{2} and \ion{C}{3} profiles in Figs. \ref{fig:lines} and \ref{fig:lines2})

Similar to the EW measurements, the C~\textsc{iii} 
$\Delta v_{90}$ are also much larger for these LRGs. 
COS-Halos have $\Delta v_{90}$ values up to 
$\sim 407$ km s$^{-1}$ 
and average line widths of $\sim 196$ km s$^{-1}$
(when smoothed to the same resolution as our QSO-LRGs spectra),
which are two to a few times smaller than the LRGs presented here. 
As with C~\textsc{iii}, 
the two LRGs also exhibit very strong \ion{H}{1} absorption 
(with \ion{H}{1}~1215 $\sim 5.6$ \AA\ and $\sim 1.7$ \AA , and column densities from the Lyman-limit flux decrement $\sim 10^{17.68}$ cm$^{-2}$ and $\sim 10^{16.99}$ cm$^{-2}$),
and the LRG\_1059+4039 H~\textsc{i} 
EW exceeds that of all COS-Halos systems. 
These two LRGs represent extrema of the CGM.  
 
We turn now to discuss the origin of this extreme CGM.
Previous studies examining systems with large 
EWs and/or large velocity widths have frequently invoked non-gravitational motions
associated with feedback processes that have ejected significant mass from the central galaxy.  This includes winds driven from 
bursts of star-formation \citep[e.g.,][]{Rubin14,Heckman17}
and outflows driven by AGN activity \citep[e.g.,][]{Tripp11}.
In our experiment, the targeted galaxies are `red and dead' with negligible
current star-formation (SF) and no signature of SF activity 
over at least the past Gyr.  Therefore, we rule out feedback from 
recent star-formation 
from LRGs as a viable origin for the observed CGM. 

We have also searched for any signatures of AGN activity from our targeted LRGs
or any neighboring galaxies. Within 2 arcmin of each LRG,
there are no sources detected in the Faint Images of the Radio Sky at Twenty-Centimeters (FIRST) survey \citep{Becker95}.  
From the Wide-field Infrared Survey Explorer \citep[WISE;][]{Wright10}, there 
are several sources in the LRG fields but all of these have
color $W1-W2 \ll 0.8$ (most have $W1-W2 \sim 0.1$), indicating
that the emission is of stellar origin \citep{Stern12}.
Together the FIRST non-detections and the WISE colors indicate 
the absence of any active AGN for galaxies near the quasar sightlines.

In addition to outflows from forming stars and AGN, gas may be 
pulled from galaxies through gravitational interactions (i.e.,\ tidal stripping). 
However, the observed velocity 
range in tidal streams 
of local interacting galaxies is typically
$\sim 100$
km s$^{-1}$ \citep{HvG96} with 
a few cases exceeding 100 km s$^{-1}$ \citep{Weilbacher03}. 
Given the observed $\sim 1000$ km s$^{-1}$ velocity spread
in the LRG-CGM, 
we consider tidal interactions to be a sub-dominant
contribution.  

Inflows of cold gas from the intergalactic medium are also disfavored because
they are predicted to have low metallicity and correspondingly much lower EWs than
what we have observed \citep{Fumagalli11}.   
Furthermore, the predicted kinematics for this infalling 
gas imply modest velocities \citep[$\approx 100~\mkms$,][]{Nelson16}   
and these cold flows are not expected to frequent massive halos
\citep{Dekel06,Keres09}.  

Absent an active source to produce the extreme CGM signatures of
our target LRGs,  we must consider alternate origins.
One possibility is that this gas is a relic 
of previous activity within the LRG (the most luminous galaxy
in the halo).  For example,
cold CGM clouds could have formed at higher redshifts, 
when LRGs were intensely forming stars or contained an AGN. 
This could provide an intragroup medium with significant
cool gas absorption \citep{Gauthier13}.  
Indeed, the CGM of QSO host galaxies contains large amounts of cold gas 
and has comparably extended and strong \ion{H}{1} and metal lines 
\citep[e.g.,][]{X13,Lau16}. 
Furthermore, the dark matter halos hosting
$z \sim 2$ quasars have estimated masses $\sim 10^{12.5} \msun$ \citep{White12}
and one predicts that a non-negligible fraction will evolve into
the halos hosting LRGs.   The obvious challenge to this scenario,
however, is that the cool gas must survive for several Gyr while
not accreting onto a galaxy or being heated as the
halo virializes.

On the point of cloud survival, idealized hydrodynamic studies of 
cool gas clouds moving through a hot and more diffuse halo
indicate that the largest sized clumps ($r > 250$\,pc) may
survive for hundreds of Myr \citep{Armillotta17}. 
Even
a population of such very large clouds, however, are
predicted to shed the majority of their mass after several Gyr
(extrapolating from the published mass loss rates).
Over time, less massive clouds would be destroyed, possibly explaining why the covering fraction of LRGs is low but certain sightlines have large amounts of cold gas detected.

The scenario we favor is that our LRG-CGM absorbers originate from gas in overlapping 
projected subhalos throughout the larger dark matter halo.
In contrast to $L^*$ galaxies, where low-mass satellites are unlikely to contribute
significantly to the observed cool CGM \citep{Tumlinson13},
the halos hosting LRGs may contain many luminous ($L \sim L^*$) satellite galaxies.
To explore this hypothesis, we have
searched for galaxies in the SDSS imaging catalog whose 
photometric redshifts  are consistent with the LRG redshift 
$z_{\rm LRG}$ (i.e., $|z_{\rm LRG} - z_{\rm phot}| < \sigma (z_{\rm phot})$), 
hereafter, neighboring galaxies (NGs).
Placing these sources at the LRG redshifts, 
we identify 3 galaxies within $R < 100$\,kpc of each
of the QSO\_1059+4039 and QSO\_1440-0156 sightlines and several additional galaxies with $R = 100 - 200$\,kpc. All 
NGs
around LRG\_1059+4039 have red colors with $u - r > 0.9$, except one (4 red, 1 blue), 
while only one of the 4 galaxies around LRG\_1440-0156 has a red color. 
Not including LRGs, 
and assuming that at $z\sim 0.5$ absolute magnitude $M_{g} \sim -20.4$ corresponds to $L_{*}$ 
galaxies \citep[][and references therein]{Skibba13}, the
$g$-band luminosities of the nearby galaxies are $\sim 0.6 - 1.5 L_{*}$. 

We offer a few additional comments on several specific galaxies.
One of the 3 galaxies around QSO\_1059+4039 is located very close to the QSO sightline
($\approx 2''$), and is not cataloged by the SDSS survey but is detected in the Panoramic Survey Telescope and Rapid Response System (Pan-STARRS) images. 
In addition, another LRG with somewhat higher stellar mass lies $\sim 200$ kpc away from the QSO\_1059+4039 sightline and is
offset by only $\approx +200$ km s$^{-1}$ from the target LRG.
Lastly, we comment that
the redMaPPer survey \citep{Rykoff14} identifies a cluster of galaxies near LRG\_1059+4039 (at distance 1.68' and with a consistent redshift) and 
no clusters within $5'$ of LRG\_1440-0156. 
In addition, these two LRGs occupy $2/3$ of the most densely populated environments in our full sample (Smailagi{\' c} et al. 2018, in preparation). 

We now explore whether the overlapping CGM of
$\sim 5$ luminous galaxies within 200\,kpc
of the sightline is sufficient to reproduce the observed
EW and $\Delta v$ values.
The estimated line-of-sight velocity dispersion for galaxies in
a dark matter halo of $10^{13.5} \msun$ is $\sigma_{1D} \approx 286 \,\mkms$.
If we randomly sample a Gaussian distribution with $\sigma_{1D}$,
the average velocity spread for 5 galaxies is $\sim 665 \, \mkms$, 
and 2 $\sigma$ range is $\sim 172 - 1158 \, \mkms $. 
If each of theses galaxies exhibits a CGM comparable to present-day
$L^*$ galaxies ($\approx 124 \, \mkms$ for COS-Halos), we recover a velocity width that is comparable to the observed velocity spreads.  

Next, consider the observed EWs. 
For every 
NG
within 200 kpc of the QSO sightline, we estimate the stellar mass using rest-frame absolute magnitudes from SDSS and the correlation between stellar mass and absolute red magnitude from \citet{LiangChen14}: $\log M_{*} = 0.14 - 0.49 M_{sdss,r}$. 
Following \citet{Bordoloi17}, we estimate the Ly$\alpha$ EWs from these
NGs
as $\log W_{HI1215} = 0.34 - 0.0026 R 
+0.286 \log M_{*}$, where $R$ is the impact parameter between galaxies and QSO. 
For the targeted LRGs only, we obtain EW 3.1\AA\ and 0.5\AA , 
which are smaller than observed.   
If we strictly add the
Ly$\alpha$ EWs from the galaxies within 200 kpc, 
we obtain 6.9\AA\ and 5.1\AA. 
When we take into account that the covering fraction for red galaxies in COS-Halos is $\sim 75$\% \citep{Tumlinson13}, the calculated EWs are $\sim 7.3$ and 3.8 \AA,
larger than the observed EWs. However, the total predicted
EWs would decrease if we allow for overlapping absorption 
if the sum is a saturated line.
Furthermore, one may expect the galaxies in dense environments to have 
smaller EWs \citep[see, e.g.,][]{Burchett17}, 
or several of the galaxies may be at unrelated redshifts. 

The overlapping subhalos CGM scenario is consistent with the multiple line components observed in our LRGs' CGM 
and
their large velocity spread. 
For example, the cold CGM could originate from outflows from blue NGs, or survive \citep[e.g.,][]{Armillotta17} or reform \citep[e.g.,][]{Thompson16} from earlier times.
 
In turn, our results support scenarios where the properties of the
CGM are more tightly linked to the underlying dark matter halo
than properties of the targeted galaxies (e.g., star-formation),
\citep[e.g.,][]{Zhu14}.
This scenario could also be tested by obtaining spectra of the putative satellite galaxies of these two LRGs
to confirm their membership and assess their properties.
The results presented here reveal that when cool CGM are present
in massive halos, the associated absorption may greatly
exceed that in lower mass systems.

\listofchanges

\end{document}